\documentclass[runningheads]{llncs}

\usepackage{graphicx}
\usepackage{booktabs}
\usepackage{amsmath}
\usepackage{hyperref}
\usepackage{xcolor}
\usepackage{microtype}
\usepackage{array}
\usepackage[normalem]{ulem}

\definecolor{myblue}{RGB}{37,165,203}
\definecolor{myred}{RGB}{175,32,67}

\newcommand{\AAcomm}[1]{}
\newcommand{\BMcomm}[1]{}

\begin{document}

\title{The Illusion of Power Capping in LLM Decode:\texorpdfstring{\\}{ }
A Phase-Aware Energy Characterisation Across Attention Architectures}

\titlerunning{The Illusion of Power Capping in LLM Decode}

\author{Bole Ma\inst{1}\orcidID{0009-0006-6536-1044}
\and Ayesha Afzal\inst{1}\orcidID{0000-0002-8632-3681} 
\and \\Jan Eitzinger\inst{1}\orcidID{0009-0000-3350-3841} \and Gerhard Wellein\inst{2}\orcidID{0000-0001-7371-3026}}
\authorrunning{B. Ma et al.}

\institute{Erlangen National High Performance Computing Center, Erlangen, Germany
\and
Friedrich-Alexander-Universität Erlangen-Nürnberg, Erlangen, Germany \\
\email{\{bole.ma, ayesha.afzal, jan.eitzinger, gerhard.wellein\}@fau.de}}
\maketitle

\begin{abstract}
Power capping is a common GPU energy lever in LLM serving,
and it appears to work: the cap is set, the GPU reports drawing less
than it, and the energy budget is met.
We show the cap does little for the phase that dominates LLM
serving: autoregressive decode.
Across four attention paradigms (GQA, MLA, Gated DeltaNet, and Mamba2)
on an NVIDIA H200 under vLLM eager execution, decode kernels wait on
HBM and the GPU idles between kernel launches, so decode draws less than
half of the GPU's 700\,W rating and throughput stays within a few
percent under every cap we set.
The cap engages only when a large batch runs a long generation.
Locking the SM clock instead lowers power
without slowing memory-bound decode and recovers up to 32\% of decode
energy at minimal throughput loss.
We identify three architecture-dependent frequency-scaling classes and
find that, once prefill is measured without prefix caching, the
recurrent architectures pay \emph{no} prefill penalty: their linear-time
mixing undercuts quadratic attention on long prompts, so their efficient
decode adds to the prefill saving and lowers total request energy from
the first output token at large batch.

\end{abstract}

\section{Introduction}
\label{sec:intro}

Data centres increasingly rely on GPU power capping to manage the
energy footprint of LLM inference~\cite{powercap,sardine}: set a
board-level watt ceiling, and the driver throttles the GPU to stay
within it, trading throughput for guaranteed power savings.
The model holds for compute-bound workloads that push the GPU near its
thermal design power (TDP).
It fails for the phase that dominates LLM serving,
\emph{autoregressive decode}, the token-by-token generation that
produces every output token after the prompt is read and accounts for
most of inference wall-clock time~\cite{mei2024dvfs,splitwise}.

Across four attention paradigms (grouped-query attention,
GQA~\cite{ainslie2023gqa}; multi-head latent attention,
MLA~\cite{deepseekv2}; Gated DeltaNet, GDN~\cite{yang2024gdn}; and the
state-space model Mamba2~\cite{mamba2}), in the memory-bound regime (single-stream decode
up to 32K output tokens, or batched decode up to ${\sim}$1{,}000 output
tokens),
power draw on NVIDIA H200 under vLLM eager execution stays at
147--292\,W on a 700\,W-TDP GPU, and
throughput moves $<$3\% across the full 280--700\,W cap range.
The SM clock does not respond to the configured limit: decode kernels
wait on HBM, eager execution leaves the GPU idle between kernel
launches, and the draw stays below the limit.
Only a large batch sustained over a long generation raises the draw
enough to reach the cap; we delimit that boundary in
Section~\ref{sec:arch-dvfs:powercap}, where the cap acts as a dynamic
underclock.
Clock settings above 1590\,MHz buy decode almost no throughput while
power rises by about 9\%.

SM clock locking lowers power
without slowing memory-bound decode and recovers up to
32\% of decode energy in the memory-bound regime at less
than 1\% throughput loss, matching the throughput of every cap setting
there at lower energy.
We characterise energy under both mechanisms for these four paradigms
(dense attention, compressed-KV, linear-recurrent, state-space), served
via vLLM on H200 SXM, with a controlled GQA$\leftrightarrow$MLA ablation via
TransMLA~\cite{transmla} on shared Minitron-4B~\cite{minitron} weights
isolating the attention mechanism.
Our contributions:
\begin{enumerate}
  \item \textbf{The power-capping illusion.}
        Power capping saves little energy in the memory-bound decode
        regimes that dominate latency-sensitive serving: the GPU draws
        too little power for even our lowest cap to change throughput
        measurably.
        We quantify the gap between configured and actual GPU behaviour
        and delimit the large-batch, long-generation boundary beyond
        which the cap does bind (Section~\ref{sec:arch-dvfs:powercap}).
  \item \textbf{SM clock locking as the better lever.}
        Energy control must act on the SM clock, not on aggregate power:
        lowering the clock cuts power without slowing a memory-bound
        step, while the cap waits for a draw that does not reach it.
        Across the memory-bound regime, a locked clock matches the
        throughput of every cap setting within 2\% at lower energy.
        We identify three architecture-dependent behavioural classes
        (batch-invariant, batch-sensitive, compute-light) and give
        deployable per-architecture clock policies.
  \item \textbf{Cross-architecture energy landscape.}
        Each novel architecture saves energy where its mechanism
        applies: recurrent models cost less in \emph{both} phases,
        19--44\% on long prompts through linear-time mixing and
        up to $2.6\times$ on long generations, while MLA saves energy
        in decode only, through its compressed KV cache, and undercuts
        GQA beyond a batch-size-dependent output length.
\end{enumerate}

\section{Background and Related Work}
\label{sec:background}

\subsection{Attention Architectures for LLM Inference}
\label{sec:background:taxonomy}

The four paradigms differ in what they store per token and how they
compute each decode step.
GQA caches full key--value (KV) pairs per group of heads and remains dominant
in mainstream open-weight transformers.
MLA caches a compressed latent KV vector (576 dims against 2{,}048) and
reconstructs KV by projection at decode time; introduced in
DeepSeek-V2~\cite{deepseekv2}, it has since propagated to GLM and
Mistral variants.
GDN replaces attention with a linear recurrence over a fixed-size state;
Qwen3.5-4B interleaves 24 GDN layers with 8 full-attention
layers~\cite{yang2024gdn}.
Mamba2 replaces attention with state-space model (SSM) layers at $O(1)$
per-step cost; Nemotron-3-Nano-4B interleaves 21 Mamba2 layers with 4
attention layers~\cite{mamba2}.
All models are ${\approx}$4B parameters.
GQA-ctrl (Minitron-4B~\cite{minitron}) carries the same GQA mechanism as
Qwen3-4B and serves as the controlled baseline for MLA: it and the
TransMLA-converted~\cite{transmla} variant, which vLLM serves through its
compressed-cache path, share base weights and differ only in attention,
so the comparison separates the attention mechanism from model weights,
vocabulary and training data.
Which kernel class dominates, and whether it is compute- or
memory-bound, decides the energy behaviour.

\subsection{GPU Clock Scaling and Power Capping}
\label{sec:background:clocks}

NVIDIA GPUs expose two static energy levers via NVML.
\emph{SM clock locking} fixes the compute-core frequency while HBM stays
at its rated clock, letting the operator set the speed--power
trade-off.
\emph{Power capping} sets a board-level ceiling and leaves clock
selection to the driver.
A cap constrains the GPU only while actual draw exceeds it: a workload
light enough to stay below the limit leaves the cap inert and the driver
at its default clock (Section~\ref{sec:arch-dvfs:powercap}).
We evaluate both, applied statically before serving.
HPC schedulers and vendor fleet managers expose power capping (NERSC surfaces only
a GPU power cap, with no clock
control)~\cite{nersc_powercap,nvidia_dcgm}, and it
makes power \emph{oversubscription} safe: hyperscalers provision less
power than the nameplate sum and cap to throttle rare
peaks~\cite{azureoversub}, a practice also applied to
LLM inference clusters~\cite{polca} and to sustainability-driven capping
at supercomputer scale~\cite{gpupowercaphpc}.
Characterisations on HPC kernels~\cite{lim2020} and
training~\cite{li2023dvfs} give a simple rule of thumb: compute-bound
kernels scale with clock, memory-bound kernels do not.
LLM inference mixes kernel types in architecture-dependent ways the rule
alone cannot predict.

\subsection{Open Questions}
\label{sec:background:gap}

Low GPU power draw during decode has been observed before.
POLCA~\cite{polca} reported the two-stage power profile from Azure fleet
telemetry, used the resulting headroom to argue for oversubscription,
and observed that a reactive power cap ``is more likely to only impact
prompt phase power draw,'' recommending frequency locking instead, on
dense transformers.
POLCA does not measure what the cap does to decode throughput and clock,
or compare the two levers head-to-head on a
throughput--energy Pareto frontier; its successor TAPAS~\cite{tapas} feeds the prefill/decode
power split into thermal- and power-aware scheduling without asking
whether the cap works for decode.
Dynamic voltage and frequency scaling (DVFS) studies~\cite{mei2024dvfs,greenllm} apply clock scaling to decode
and report savings against default-governor baselines rather than
against power capping.
We test POLCA's observation directly, answering \emph{is power
capping effective for memory-bound LLM decode?} with a direct
cap-vs-clock Pareto comparison across four attention paradigms,
including the compressed-KV and recurrent designs POLCA did not study,
and delimiting the corner where the draw rises and the cap re-engages.

Inference studies focus on throughput and
latency~\cite{vllm,flashattn,flashattn3}; those reporting
energy or applying DVFS stay within standard GQA/MHA
transformers~\cite{polca,mei2024dvfs,greenllm}, and none of them
examines MLA's compressed KV path, linear-attention hybrids such
as GDN, or SSM models such as Mamba2, whose different kernel types may
alter the DVFS response.

\section{Experimental Setup}
\label{sec:methodology}

\subsection{Deployment Baseline and Hardware}
\label{sec:methodology:baseline}

Every model is served via vLLM~\cite{vllm} in BF16 from its HuggingFace
checkpoint, with no custom kernels.
All models run in vLLM's eager mode, without CUDA-graph capture.
We run on a single NVIDIA H200 SXM (HBM3e, 4.8\,TB/s bandwidth,
989\,TFLOPS BF16 dense peak, 700\,W TDP) with NVIDIA driver 590.48.
The single-card setup mirrors the \emph{decode pool} model of
disaggregated serving~\cite{splitwise,distserve}, where each
decode-pool card sees a decode-only workload.
Energy is the difference of the board's NVML energy counter over each
run, and average power comes from NVML power readings sampled every
50\,ms.
The counter updates at about 10\,Hz, so every run we report lasts at
least two seconds; over decode runs, the integrated power samples agree
with the counter to within 2\%.
Prefill is measured with vLLM's automatic prefix caching \emph{disabled}
and a distinct random-token prompt per sequence and per repetition.
With the engine defaults, repeated identical prompts
turn every post-warmup ``prefill'' into a KV-cache hit, understating
prefill energy by more than an order of magnitude.
Because vLLM disables prefix caching for the recurrent hybrids, the
understatement hits \emph{only} the attention-based models, which biases
any cross-architecture comparison.
Nsight Compute (NCU) profiles one vLLM engine step per model in the same
configuration (FlashAttention-3, eager execution, unlocked clocks,
driver 595.91.07), under the 600\,W power limit that our computing
centre sets on its H200 nodes in place of the 700\,W board default,
counting each kernel's executed operations, DRAM bytes and SM clock.

\subsection{Experimental Design}
\label{sec:methodology:design}

Our primary metric is \emph{energy per token} (mJ/tok), measured
separately for prefill and decode.
For prefill, the denominator is the number of \emph{input} tokens
processed (i.e.\ prompt length~$\times$~batch size); for decode it is the
number of \emph{output} tokens generated.
Throughput is generated tokens per second; throughput loss and energy
saving are the relative drop in throughput and in energy per token
against the highest setting of the same sweep (1980\,MHz requested
clock, 700\,W cap).
We sweep five SM clock levels from 390 to 1980\,MHz (with HBM held at
rated speed), five power cap levels from 280 to 700\,W, batch sizes
from 1 to 32, and lengths from 1K to 64K tokens.
A prefill run processes a prompt of that many \emph{input tokens}; a
decode run starts from a one-token prompt and generates that many
\emph{output tokens}, so its KV cache grows to that length.
We label every run by these counts.
The lowest cap sits at 40\% of the 700\,W TDP; the full range brackets
the operating envelope.
Warmup iterations and repetitions fall with run length, from three
warmups and ten runs (the first discarded) at 1K tokens to a single run
with at most one warmup for decode at 16K output tokens and above, where
one run averages power over thousands of decode steps; we report medians
with $\pm$1 standard deviation bands.

\section{The Hardware Substrate}
\label{sec:substrate}

\begin{figure}[t!]
  \centering
  \includegraphics[width=0.49\linewidth]{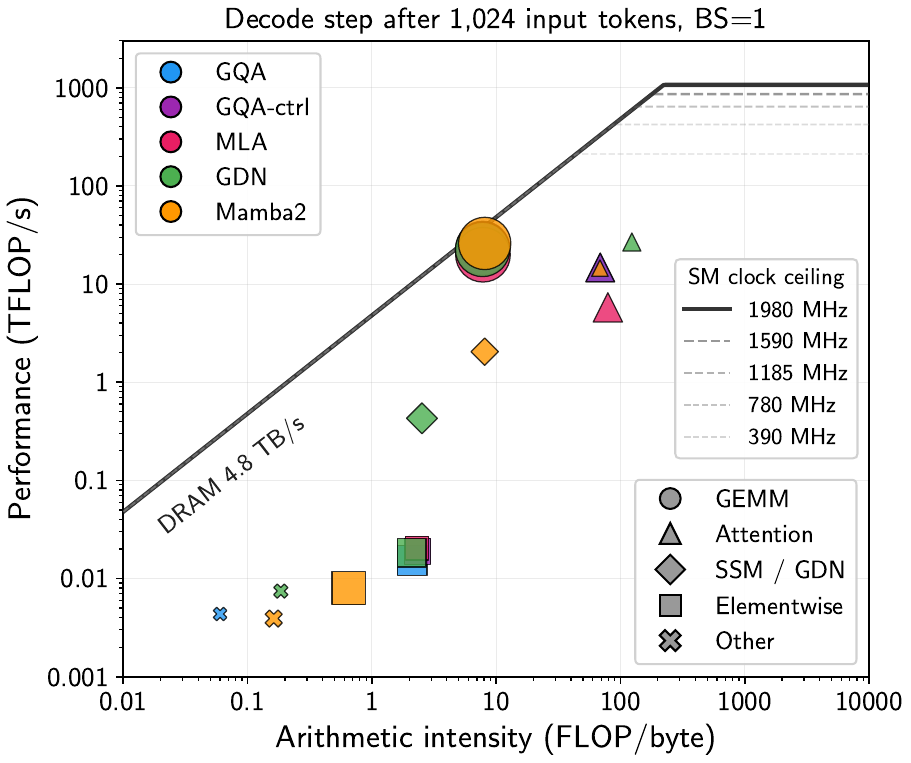}
  \hfill
  \includegraphics[width=0.49\linewidth]{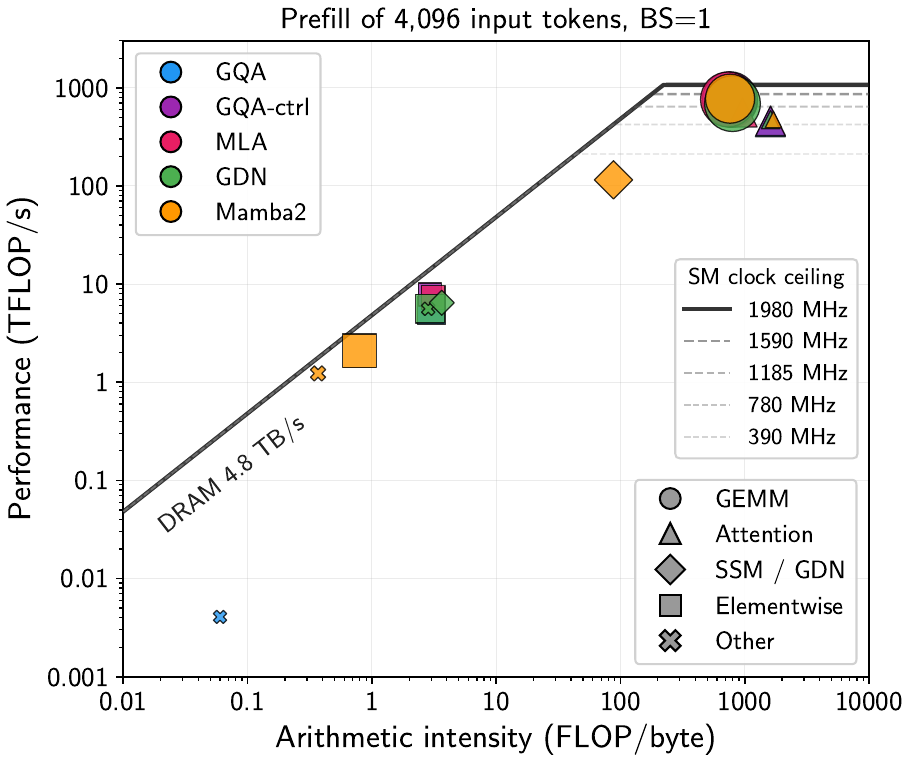}
  \caption{H200 roofline of one vLLM engine step per model, profiled with
           unlocked clocks under the computing centre's 600\,W power limit: one
           decode step after a prompt of 1{,}024 input tokens (left) and the
           prefill of 4{,}096 input tokens (right), both at BS\,=\,1, each marker one kernel class at its
           counted FLOPs per DRAM byte and achieved FLOP rate, sized by its
           share of the step's GPU time, with the compute ceiling drawn at
           each swept SM clock.}
  \label{fig:roofline}
\end{figure}

The H200's roofline ridge sits at $\approx$206\,FLOPs/byte: kernels
below this threshold are waiting on memory and do not benefit from faster
compute; kernels above it are compute-limited and scale with clock.
Prefill processes all tokens in parallel via large GEMMs, which run
compute-bound.
Decode at BS\,=\,1 reduces each multiply to a matrix--vector
operation, which runs memory-bound; the decode kernels labelled GEMM in
Figure~\ref{fig:roofline} compute these matrix--vector products.
NCU counts executed tensor operations, which include the 8-wide tile
that pads each batch-1 product, so these kernels plot near
8\,FLOP/byte instead of the 1\,FLOP/byte of the useful work.
In prefill, the GEMMs and attention reach the compute roof while the
GDN and Mamba2 recurrences stay on the bandwidth side.
GEMMs take the largest share of GPU time in both phases, 57--75\% of a
prefill step and 65--72\% of a decode step, so their place on the
roofline decides how each phase responds to the SM clock.
In prefill they sit at the compute ceiling of the clock they run at, so
a lower clock slows prefill down; in decode they sit on the bandwidth
slope, where they wait on memory at any clock.
This contrast gives decode the clock headroom that
Section~\ref{sec:arch-dvfs} exploits.
With clocks unlocked, the driver runs each kernel at its own SM clock:
the GEMMs run lowest, near 1.83\,GHz in decode and 1.5--1.6\,GHz in
prefill under the 600\,W limit, while light kernels run close to
1.98\,GHz, so Figure~\ref{fig:roofline} draws the compute ceiling at each
swept clock.

\subsection{Decode Kernels Wait on Memory}
\label{sec:substrate:decode}

Decode kernels spend most of their time loading model weights from HBM,
and the tensor cores run at a small fraction of their peak throughout a
decode step.
Under eager execution the GPU also idles between kernels while the host
launches the next one: at the default clock and 1K output tokens, NVML reports
a kernel running for less than half of the time, for every architecture
and batch size.
Neither the memory-bound kernels nor the launch gaps depend on the SM
clock.
A lower clock slows each kernel, but the extra time comes out of the
idle gaps: the share of time a kernel runs rises while throughput holds, so up to 4K output tokens
underclocking saves energy for every architecture and batch size we
tested (Section~\ref{sec:arch-dvfs:spectrum}).
This holds only down to a floor that depends on architecture and batch
size.
At 1K output tokens and BS\,=\,1, GQA and GQA-ctrl hold throughput down to
${\approx}$780\,MHz and dip at 390\,MHz, while MLA, GDN and Mamba2 hold
it even at 390\,MHz; larger batches and longer generations raise the floor
(Table~\ref{tab:policy}).
Below the floor, the slower kernels use up all the idle time and
underclocking costs throughput.

\subsection{Batch Size and Clock Policy}
\label{sec:substrate:batch}

Batching has a far larger effect than any DVFS or architecture choice:
at 1K output tokens, increasing BS from 1 to 32 reduces energy-per-token by over
20$\times$
by amortising the cost of loading weights.
But batching also shifts the compute/memory balance in
architecture-dependent ways, revealing three DVFS classes
(Figure~\ref{fig:dvfs-heatmaps-all}).
At 1K output tokens, \emph{batch-invariant} architectures (GQA, GQA-ctrl)
keep a low clock at every batch size.
\emph{Batch-sensitive} architectures (MLA, Mamba2) contain enough
additional per-step work (extra projection GEMMs for MLA, SSM scan
compute for Mamba2) that large batches make them sensitive to the SM
clock; the energy-optimal clock must rise with batch size.
\emph{Compute-light} GDN stays within the 5\% throughput-loss budget at
390\,MHz even at BS\,=\,32 (Table~\ref{tab:policy}) and tolerates the
most aggressive underclocking at every batch size and length we tested.
Long generations reorder the classes: at 16K output tokens and
BS\,$\geq$\,16,
GQA and GQA-ctrl need 1590\,MHz to stay within the 5\% budget while MLA
stays at 1185\,MHz.
Table~\ref{tab:policy} turns the classes into a static per-architecture
clock policy: batch-sensitive architectures must raise their clock once
the serving engine fills large batches.

The power that does not follow the SM clock caps what DVFS can deliver:
a 5$\times$ clock reduction yields only ${\sim}$1.5$\times$ less power,
since DVFS moves the dynamic component alone.
That component scales as ${\sim}V^2 f$, and a lower SM clock also lowers
the coupled core voltage.

\begin{table}[t]
  \centering
  \caption{Decode SM clock policy at 1K output tokens (5\%
    throughput-loss budget), applied before serving.
    GQA-ctrl is the MLA ablation control: same GQA attention as
    Qwen3-4B, shared weights with the MLA variant.}
  \label{tab:policy}
  \small
  \begin{tabular}{@{}l lll l@{}}
    \toprule
    \textbf{Arch} & \textbf{BS=1} & \textbf{BS=16} & \textbf{BS=32} & \textbf{Class} \\
    \midrule
    GQA      & 780\,(--24\%)  & 780\,(--26\%)  & 780\,(--26\%)  & Batch-invariant \\
    GQA-ctrl & 780\,(--25\%)  & 1185\,(--21\%) & 780\,(--27\%)  & Batch-invariant \\
    MLA      & 780\,(--24\%)  & 1185\,(--24\%) & 1185\,(--27\%) & Batch-sensitive \\
    GDN      & 390\,(--30\%)  & 390\,(--29\%)  & 390\,(--32\%)  & Compute-light \\
    Mamba2   & 390\,(--25\%)  & 780\,(--29\%)  & 1185\,(--28\%) & Batch-sensitive \\
    \bottomrule
  \end{tabular}
\end{table}

\section{Power Capping vs.\ Clock Locking by Architecture}
\label{sec:arch-dvfs}

\subsection{Clock Locking}
\label{sec:arch-dvfs:spectrum}

\begin{figure}[t!]
  \centering
  \includegraphics[width=\linewidth]{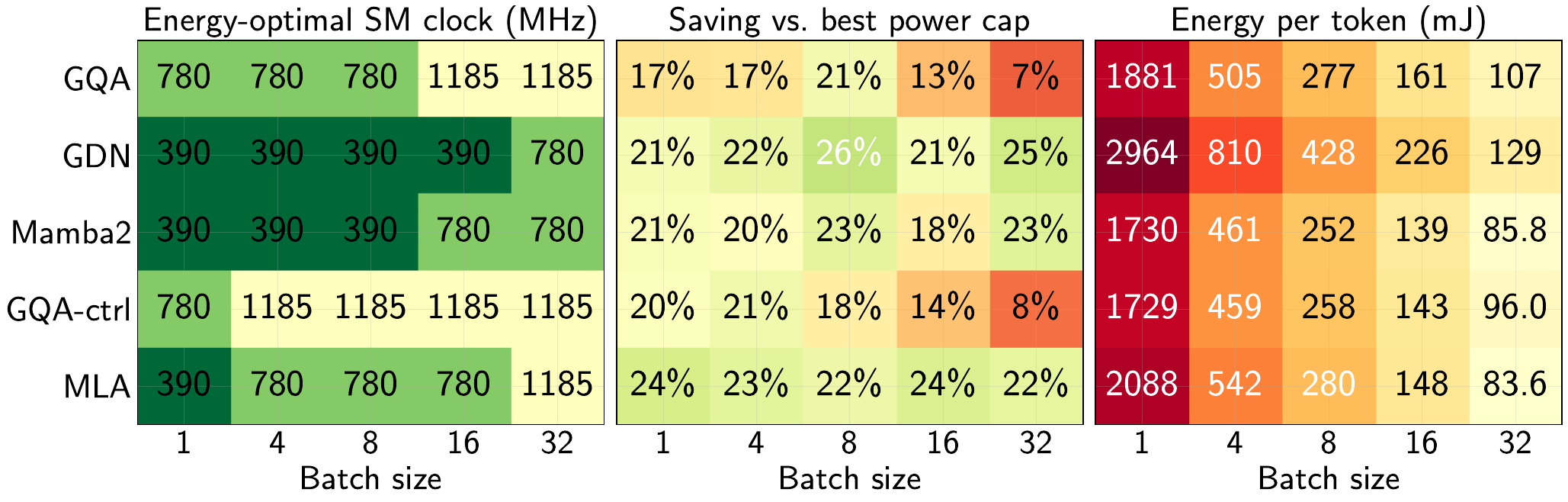}\\[4pt]
  \includegraphics[width=\linewidth]{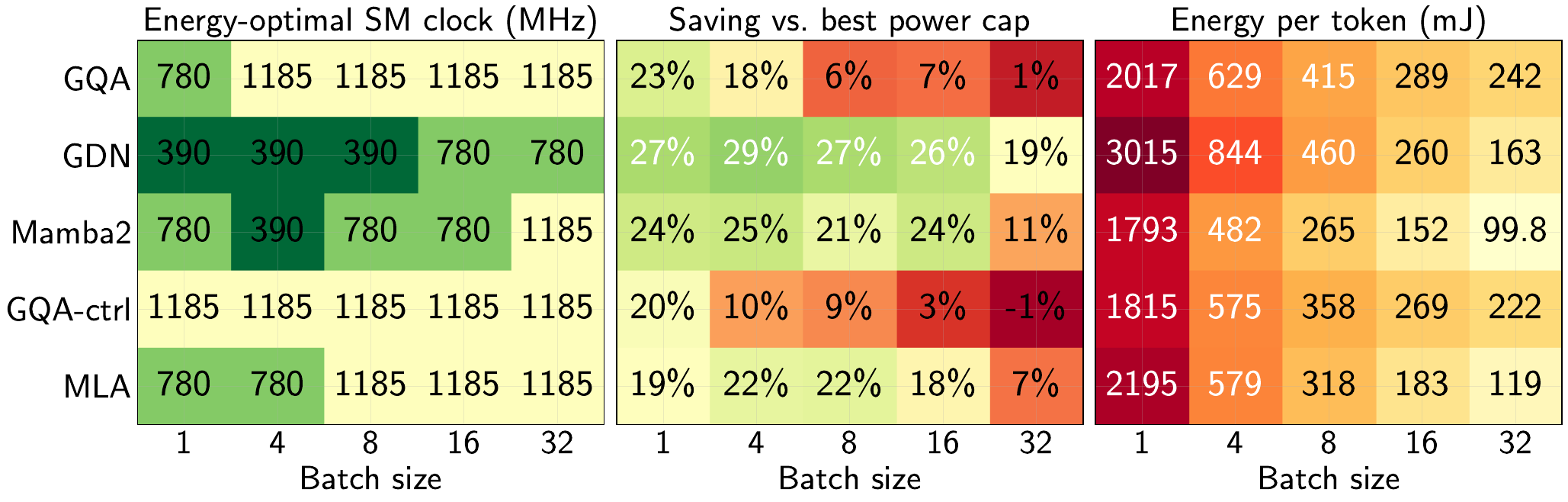}
\caption{Decode DVFS heatmaps: energy-optimal SM clock (left), its
         energy saving over the best power cap (centre), and energy per
         token at that clock (right), at 4K (top) and 16K (bottom)
         output tokens.}
  \label{fig:dvfs-heatmaps-all}
\end{figure}

Up to 4K output tokens, every architecture saves 20--32\% of its decode energy
by underclocking, relative to the 1980\,MHz setting.
Figure~\ref{fig:dvfs-heatmaps-all} maps the energy-optimal clock and its
saving over the best power cap, which shrinks where a large batch lets
the cap bind.
At 780\,MHz and 1K output tokens the saving holds at every batch size with
about 1\% throughput loss or less, and GDN benefits most; only GQA-ctrl
at BS\,=\,32 loses more (4\%).
With 16K or more output tokens at large batch, GQA and GQA-ctrl turn sensitive
to the SM clock: within the 5\% budget their clock rises to 1590\,MHz
and the saving falls below 10\%, while MLA, Mamba2 and GDN keep most of theirs.

Clock settings above 1590\,MHz buy decode almost no throughput.
The clock lock (\texttt{nvidia-smi --lock-gpu-clocks}) honours every
setting up to 1590\,MHz; across all four paradigms, the GQA-ctrl control,
1K--64K output tokens and batch sizes 1--32, the median throughput
difference between the 1590 and 1980\,MHz settings is $<$0.1\%, and
half the configurations invert (1590\,MHz faster).
Memory and kernel launches set the pace, so the extra cycles add about
9\% power and no throughput.

\begin{figure}[t!]
  \centering
  \includegraphics[width=\linewidth]{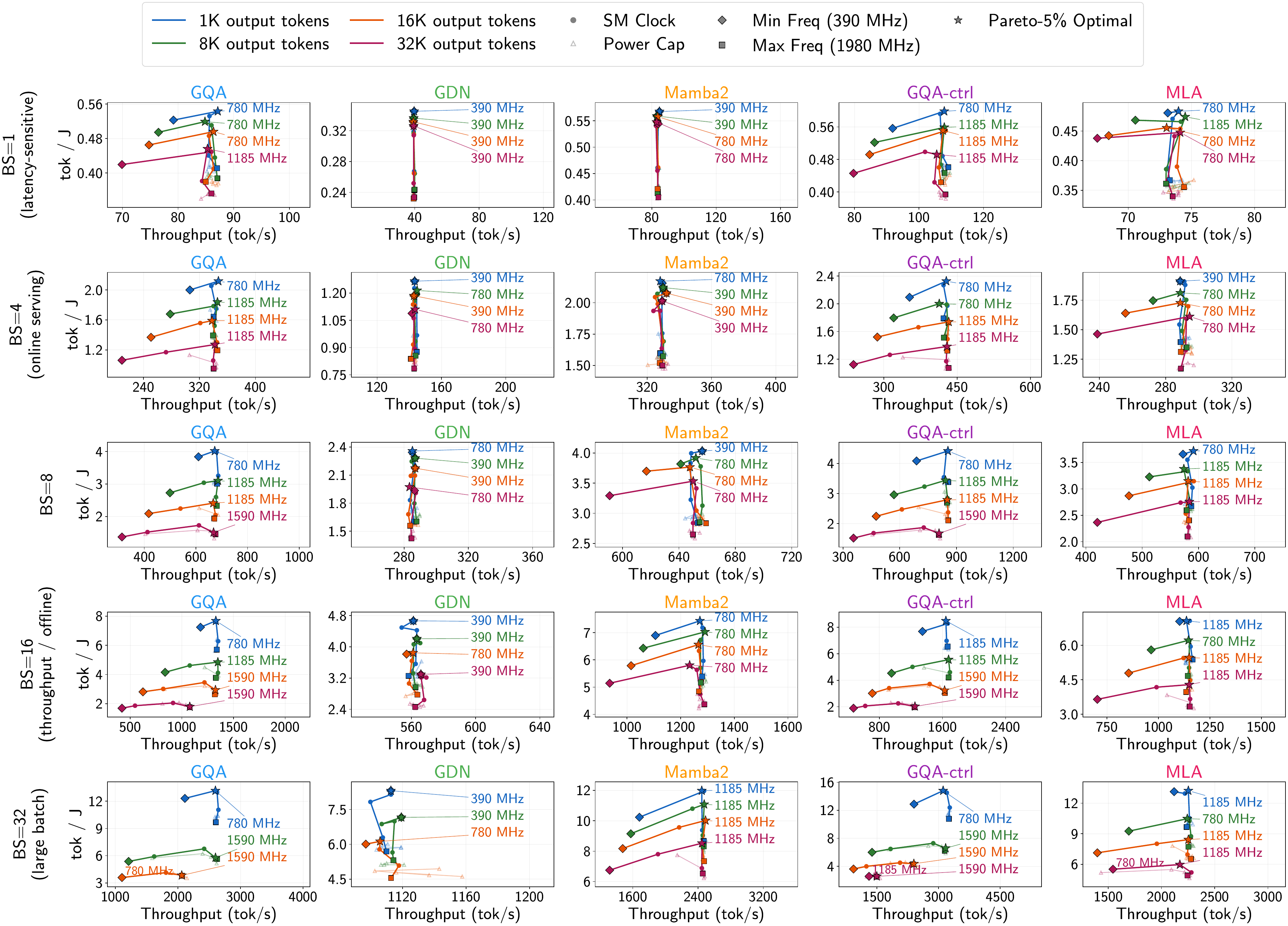}
  \caption{Decode DVFS Pareto frontier: throughput against tokens per joule
           for every SM-clock and power-cap setting, one curve per number of
           output tokens.}
  \label{fig:pareto-decode}
\end{figure}

The Pareto frontiers (Figure~\ref{fig:pareto-decode}) plot throughput
against tokens per joule for every clock and power-cap setting.
In the memory-bound regime, a locked clock matches the throughput of
every cap setting within 2\% at lower energy, a quarter lower in the
median case, while the cap settings collapse into a single cluster.
At large batch and long generation the cap settings spread out and act
as a dynamic underclock, with energy comparable to the best locked
clock.
Static clock locking costs one \texttt{nvidia-smi} call at job start.

\subsection{Why Power Capping Fails for Memory-bound Decode}
\label{sec:arch-dvfs:powercap}

\paragraph{Configured cap vs.\ actual power draw.}
Table~\ref{tab:powercap-actual} exposes the gap between what an operator
\emph{configures} and what the GPU \emph{does}.
Under every cap from 280\,W to 700\,W the NVML-reported power draw stays
within 150--215\,W and the SM clock holds ${\approx}$1830\,MHz in 13 of
15 cells; the two exceptions (GQA at 420\,W, GDN at 600\,W) come from
run-to-run firmware clock variation, which lowers power a little and
leaves throughput unchanged.
A cap only lowers power, and at 150--215\,W this decode stays below even
our lowest cap, so the cap has nothing to act on.
Throughput varies by less than 2\% across caps.

\paragraph{Where the memory-bound regime ends.}
The cap stays inert across the memory-bound regime: single-stream decode
up to 32K output tokens (150--282\,W) and batched decode up to
${\sim}$1{,}000 output tokens ($\le$292\,W even at BS\,=\,32).
The regime ends where a large batch runs a long generation: each step
streams a growing KV cache, decode turns sensitive to the SM clock, and
the draw climbs past the cap.
The attention models draw the most, approaching the 700\,W rating, and
a 280\,W cap cuts their throughput by up to 44\%, while GDN keeps its
throughput.
There the cap lowers the SM clock much as a lock does, and the best cap
reaches energy comparable to the best locked clock.

\begin{table}[t]
  \centering
  \caption{Power cap vs.\ actual GPU behaviour during decode (BS\,=\,1,
    1K output tokens, median over 9 reps).
    A 2.5$\times$ range in the configured cap leaves actual clock and
    power unchanged: the cap does not engage here.}
  \label{tab:powercap-actual}
  \small
  \begin{tabular}{@{}l rrr rrr@{}}
    \toprule
    & \multicolumn{3}{c}{\textbf{SM clock (MHz)}}
    & \multicolumn{3}{c}{\textbf{Power (W)}} \\
    \cmidrule(lr){2-4}\cmidrule(lr){5-7}
    \textbf{Cap (W)} & \textbf{GQA} & \textbf{GDN} & \textbf{MLA}
                      & \textbf{GQA} & \textbf{GDN} & \textbf{MLA} \\
    \midrule
    280  & 1830 & 1830 & 1830 & 211 & 162 & 200 \\
    420  & 1590 & 1830 & 1830 & 192 & 168 & 202 \\
    500  & 1830 & 1830 & 1830 & 206 & 167 & 201 \\
    600  & 1830 & 1590 & 1830 & 207 & 150 & 202 \\
    700  & 1830 & 1830 & 1830 & 215 & 162 & 202 \\
    \bottomrule
  \end{tabular}
\end{table}

\section{The Compressed and Recurrent Architecture Crossover}
\label{sec:crossover}

Novel attention replacements---MLA's compressed KV cache, GDN's linear
recurrence, Mamba2's SSM---are frequently assumed to trade a prefill
penalty for decode efficiency.
Measured without prefix caching, the recurrent architectures show no
such trade-off and cost less in \emph{both} phases; only MLA pays a
small prefill tax, caused by its decompression kernels.

\subsection{Prefill: Recurrent Models Cost Less}
\label{sec:crossover:prefill-penalty}

Measured with prefix caching disabled and unique prompts
(Section~\ref{sec:methodology:design}), prefill energy per token at
BS\,=\,32 and 16K input tokens spans a narrow band across all five
architectures: 10.1 (Mamba2) to 13.9\,mJ/tok (GQA).
The recurrent models prefill \emph{cheapest}: Mamba2 and GDN undercut
the attention models by 19--28\% at 16K and by 35--44\%
at 32K input tokens, as first-order FLOP analysis predicts, since
linear-time state mixing avoids the $O(N^2)$ attention term in the
prompt length $N$.
Engine defaults invert this picture: with prefix caching enabled,
repeated benchmark prompts reduce transformer ``prefill'' to a cache
hit ($<$0.5\,mJ/tok) while recurrent models, for which vLLM disables
prefix caching, run the full computation, which manufactures an
order-of-magnitude recurrent penalty.
Checkpointing the recurrent state at chosen token positions extends
prefix caching to the hybrids~\cite{marconi}, a lever our
distinct-prompt measurements leave out: with shared prefixes, both
families could skip repeated prefill work.
Unfused GDN/Mamba2 kernels
(Section~\ref{sec:discussion:limits}) leave further headroom.
MLA pays a small prefill tax over GQA-ctrl under identical weights at
16--32K input tokens.
vLLM's MLA prefill rebuilds full key and value heads from the latent
cache with an up-projection GEMM and concatenations; in an NCU profile
of the prefill of 4{,}096 input tokens, these extra GEMM and concatenation
kernels account for most of MLA's 6\% longer step.

\subsection{The Decode Payoff}
\label{sec:crossover:decode-payoff}

Mamba2's per-step decode cost grows little as the generation grows, since
only 4 of its 25 mixing layers keep a KV cache, which gives it more than
a $2\times$ energy
advantage over GQA at large batch and 16K output tokens
(Figure~\ref{fig:dvfs-heatmaps-all}, bottom).
Growing batch makes SSM scan compute significant and raises the
energy-optimal clock, placing Mamba2 in the batch-sensitive class.

MLA crosses over more gradually.
For short generations, compressing the KV cache saves nothing: weight loading
dominates HBM traffic, and MLA's decode path adds work, leaving MLA
12--25\% \emph{worse} than GQA-ctrl.
vLLM's MLA decode never rebuilds full heads; it passes queries and
outputs through absorbed projection weights instead, and in an NCU
profile of one decode step after a prompt of 1{,}024 input tokens these extra GEMMs
account for most of MLA's 19\% longer step.
As the generation grows, KV cache traffic grows linearly, steeply for GQA-ctrl
and gently for MLA's compressed latent, until MLA undercuts GQA-ctrl.
Larger batches move this crossover to shorter generations: at
BS\,=\,32 MLA undercuts GQA-ctrl from 4K output tokens on, and at
BS\,=\,1 only at the longest generation we tested.

\subsection{Total Request Energy}
\label{sec:crossover:total-energy}

\begin{figure}[t!]
  \centering
  \includegraphics[width=0.7\linewidth]{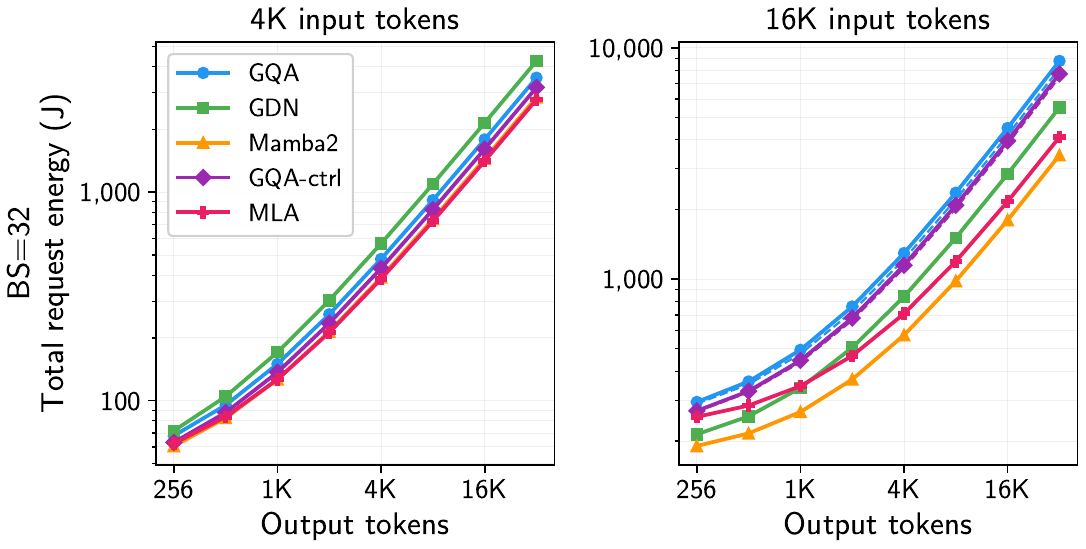}
  \caption{Total request energy against the number of output tokens
           after a prompt of 4K or 16K input tokens, with prefill measured
           without prefix caching and decode energy per token from runs
           that generate as many output tokens as the prompt has input tokens, at the Pareto-5\% clock (solid) or the
           minimum-energy clock (dashed), at BS\,=\,32.}
  \label{fig:total-request-energy}
\end{figure}

Figure~\ref{fig:total-request-energy} compares the energy of a
\emph{complete request} across architectures at BS\,=\,32.
After a prompt of 16K input tokens, Mamba2 costs least from the \emph{first} output token
on, with MLA close behind through cheaper decode.
Mamba2 saves in both phases: it spends about a quarter less per prefill
token than GQA (Section~\ref{sec:crossover:prefill-penalty}) and
$2.6\times$ less per decode token at the Pareto-optimal clock, so its
decode saving has no prefill premium to amortise.

\section{Discussion}
\label{sec:discussion}

\subsection{Implications for Data-Centre Power Management}
\label{sec:discussion:datacenter}

Cluster schedulers set per-GPU or per-node power limits to meet
facility-level budgets~\cite{powercap}, which works for workloads near
TDP: training runs, dense HPC kernels, large-batch prefill.
In our measurements, decode-dominated serving (agentic multi-turn,
streaming, latency-sensitive chat) holds the GPU in a low-power,
memory-bound state where a 280\,W cap delivers little energy reduction.

Static SM clock locking replaces the cap.
Under disaggregated serving (Splitwise~\cite{splitwise},
DistServe~\cite{distserve}) each pool locks at the energy-optimal clock for its phase
with no dynamic switching; under colocated serving a global 780\,MHz
decode clock saves 21--32\% of decode energy at 1K output tokens.

\subsection{Limitations}
\label{sec:discussion:limits}

The software stack matters as much as the architecture.
vLLM serves GDN and Mamba2 through unfused kernels, so their measured
efficiency understates what fused kernels~\cite{fla,cula} deliver, and
engine-default prefix caching fabricates an order-of-magnitude prefill
advantage for attention models
(Section~\ref{sec:methodology:design}).
Our measurements cover a single GPU (no multi-GPU communication
energy), a single framework (TensorRT-LLM may differ), and dense models
only (MoE routing may interact with DVFS differently).
Our attribution also assumes components that run apart: prefill and
decode, kernel classes, and the host-side gaps between kernels.
Eager execution leaves the GPU idle in these gaps, and part of the clock
slack we measure comes from them.
Execution paths that remove the gaps, such as CUDA-graph capture or
megakernels that run a whole forward pass as one persistent kernel,
raise the draw toward the cap, change the clock response, and couple
the components, so a per-phase or per-kernel attribution like ours
becomes hard to make.

Repeating the core decode characterisation on NVIDIA driver~595.71.05,
succeeding the 590.48 of our primary measurements, reproduces both
central findings: the power-cap throughput spread stays negligible, and the same high-draw corner appears
at large batch and long generation.
The illusion persists on driver 595.71.05; other GPU generations and
alternative lock mechanisms may behave differently.

Tensor-parallel decode raises per-GPU utilisation and could in
principle push draw above a cap.
Our batch-size sweep bounds the concern: at BS$\,=$\,32 with
short-to-moderate output, per-GPU power peaks at 292\,W across all five
architectures, 4\% above our lowest cap, where the throughput effect
stays under 3\%.

\section{Conclusion}
\label{sec:conclusion}

Under vLLM eager execution, power capping, a common data-centre
energy lever, saves little energy in memory-bound LLM decode: the GPU
draws far below its rating, no cap we apply changes throughput by more
than a few percent, and the SM clock does not respond to the cap.
Only long generation at large batch raises the draw enough for the
cap to act.
Clock settings above ${\approx}$1590\,MHz buy decode almost no
throughput while power keeps rising, so the configured power limit misrepresents
what the GPU does and the top clock settings waste energy.

Wherever decode is memory-bound, the SM clock is the lever to tune:
lowering it cuts power without slowing decode, while a power cap
waits for a draw that never comes.
Clock locking recovers up to 32\%
of decode energy at negligible throughput loss and, across the
memory-bound regime, matches the throughput of every cap setting at
lower energy; the energy-optimal clock depends on architecture, batch
size and output length, and our three behavioural classes and policy
table capture it for short generations.

Across architectures, the recurrent models pay no prefill premium:
linear-time prefill undercuts quadratic attention on long prompts, so
their efficient decode adds to that saving and lowers total request
energy relative to GQA from the first output token at large batch
sizes, while only MLA pays a small prefill tax for KV compression that
saves energy in decode.
Future work should characterise fused execution, from CUDA graphs to
megakernels, which removes the host-side gaps between kernels, raises
the draw, and couples the components we separate here
(Section~\ref{sec:discussion:limits}), as well as other GPU generations; whether fused recurrent kernels widen the
recurrent advantage; whether KV quantisation shifts the MLA crossover;
and whether adaptive frequency scaling can outperform static clock
locking at runtime.
\begin{credits}
\subsubsection{\ackname} This work has been funded by the Free State of Bavaria in the DSgenAI project (Grant Nr.: RMF-SG20-3410-2-18-4). The authors gratefully acknowledge the scientific support and HPC resources provided by the Erlangen National High Performance Computing Center (NHR@FAU) of the Friedrich-Alexander-Universität Erlangen-Nürnberg (FAU). The hardware is funded by the German Research Foundation (DFG).
\subsubsection{\discintname} The authors have no competing interests to declare that are relevant to the content of this article.
\subsubsection{Use of AI Tools.} The authors used Anthropic's Claude Opus as an interactive tool for drafting, consistency checking, polishing, and review, and take full responsibility for the content of this paper.
\end{credits}

\bibliographystyle{splncs04}
\bibliography{CP049}

\newpage

\end{document}